\documentstyle[sprocl]{article}

\def\Journal#1#2#3#4{{#1} {\bf #2}, #3 (#4)}

\def\NPB{{\em Nucl. Phys.} B}

\def\JHEP{\em J. High Energy Phys.}

\def\be{\begin{equation}}
\def\ee{\end{equation}}
\def\bea{\begin{eqnarray}}
\def\eea{\end{eqnarray}}
\begin{document}

\title{SOFT GLUON RESUMMATION FOR SHAPE VARIABLE DISTRIBUTIONS IN DIS}

\author{M. DASGUPTA}

\address{Dipartimento di Fisica, Universita di Milano Bicocca and
  INFN, Sezione di Milano, Italy}


\maketitle\abstracts{We discuss the procedure to resum large
  logarithms to all orders in the differential distributions for DIS
  event shape variables $\frac{\tau}{\sigma}\frac{d\sigma}{d\tau}$. We
  describe results for two variants of the thrust variable, both
  defined wrt the boson axis in Breit frame but with
  differing normalisations.}

\section{Introduction}
Event shape variables in $e^{+}e^{-}$ reactions 
have long been the subject of much attention
from theorists and experimentalists proving popular tools for the
extraction of $\alpha_s$ as well as for the study of non-perturbative
(power) corrections. In the last few years such
attention has also been focussed on similar variables in DIS with
experimental studies being carried out by both the H1 \cite{Rabbertz}
and Zeus collaborations. \cite{Wollmer} In this regard one area of investigation involves the
study of differential distributions in shape variables $\tau$ which have
the perturbative expansion
\begin{equation}
\label{one}
\frac{1}{\sigma}\frac{d \sigma}{d\tau} = A_1(\tau,x)\alpha_s +
A_2(\tau,x)\alpha_s^2+\cdots 
\end{equation}

While, in the above equation, the coefficient $A_1$ is obtainable
through an analytical calculation, $A_2$ can be provided by NLO Monte Carlo
programs DISENT \cite{CatSey} and DISASTER++. \cite{Grau} The variable 
$x$ is the standard Bjorken variable.   
At present, experimental studies are confined to comparing the above
truncated expansion to data. \cite{Rabbertz} The findings are that while such a comparison is
viable for larger values of $\tau$ it gets progressively meaningless as
we go to the small $\tau$ region. The perturbative results appear to
diverge at small $\tau$ instead of turning over as indicated by the data.

This problem is familiar from
$e^{+}e^{-}$ distributions. Essentially the perturbative coefficients
  above have a leading behaviour (in the small $\tau$ region)
\begin{equation}
A_n (\tau,x) \sim \frac{1}{\tau}\ln^{2n-1} \left ( \frac{1}{\tau} \right )+\cdots
\end{equation}
At small $\tau$ the smallness of $\alpha_s$ is more than compensated by
the large $\ln(1/\tau)$ terms which accounts for the divergent behaviour
of fixed order results in that regime. 
In what follows we shall also refer to the integrated {\it{shape cross 
    section}} $$R(\tau) =
\int_{0}^{\tau}\frac{1}{\sigma}\frac{d\sigma}{d\tau'}d\tau'$$ for which for every
power of $\alpha_s$ there are up to two powers of $\ln(1/\tau)$
(henceforth denoted by $L$). The differential distribution can be
obtained from $R$ by straightforward differentiation.  

The leading $(\alpha_s L^2)^n$ behaviour in the perturbative
prediction for $R$ is transparently
associated with $n$ soft gluon emission and the absence of complete
Bloch-Nordsieck cancellations. Such terms are free from any $x$
dependence since soft gluon emission does not change significantly the 
momentum fraction of the incoming projectile. However there are other
sources of large logarithms apart from purely soft emission. These
include running coupling effects and hard collinear emissions.
It turns out that hard collinear emissions on the incoming leg in DIS
are responsible for terms starting at $(\alpha_s L)^n$ (single
logarithmic level)with coefficents which are $x$ dependent.  

Clearly in order to obtain meaningful perturbative results, one is
required to try and resum these large logarithms to all orders in
perturbation theory. In this regard the only new feature of the DIS resummation in
comparison to $e^{+}e^{-}$ shape variable resummation is the above mentioned $x$ dependent single 
  logs. Their resummation is possible since they are of 
  leading log DGLAP type and lead to a change of scale in the parton
  distribution, or the emergence of the factor $\frac{q(x,\tau Q^2)}{q(x,Q^2)}$
in the final result. \cite{ADS} The other aspects of the resummation
are mainly familiar from experience with $e^{+}e^{-}$ variables, though the details
differ from variable to variable. Note that in writing the factor
$q(x,\tau Q^2)$ we have
also implicitly included the contribution from incoming gluons. 

\section{Definitions}
We consider below the following two definitions of the thrust variable in
the Breit current hemisphere:
\begin{equation}
\tau_Q = 1-T_Q = 1-\frac{2}{Q} \sum_{i\in H_c} |\vec{P}_i.\hat{n}|
\end{equation}
\begin{equation}
\tau_E = 1-T_E = 1-\frac{\sum_{i \in H_c} {|\vec{P}_i.\hat{n}|}}{\sum_{i 
    \in H_c}|\vec{P}_i|}
\end{equation}
where the unit vector $\hat{n}$ denotes the photon direction in the
Breit frame and the sum is over all particles in the current
hemisphere $H_c$.
The definitions differ only in normalisation but we find that this
significantly affects the resummation both in procedure and results. \cite{ADS}
The latter definition, involving
normalisation to the current hemisphere energy, is only infrared safe
provided one imposes a minimum energy cut-off in $H_c$ .

\section{Results and Conclusions}
The resummed result for the contribution to $F_2$ from events with $1-T<\tau \ll 1$ (where $\tau$ is 
now used to denote $\tau_Q$ or $\tau_E$ ) is given by 
\begin{equation} 
R(x,Q^2,\tau) = x\left [\sum_{q,\bar{q}} e_q^2 q(x,\tau Q^2)
    +C_{q}(x)+C_{g}(x)\right ] \Sigma(\alpha_s,L) 
\end{equation}
Here $C_q$ and $C_g$ are ${\mathcal{O}}(\alpha_s)$ constant pieces from
incoming quark and gluon sectors respectively and they are different
for $\tau_Q$ and $\tau_E$. 
Additionally one has the form factor
\begin{equation}
\Sigma(\alpha_s,L) = \exp [ L g_1(\alpha_s L) +g_2(\alpha_sL)+\alpha_s 
g_3(\alpha_s L)+\cdots]
\end{equation}
in which we control all terms of the type $\alpha_s^n L^{n+1}$ and
$\alpha_s^n L^n$. In other words the functions $g_1$ and $g_2$ are
explicitly computed \cite{ADS} while $g_3$ is unknown. All these
functions also depend on the variable under consideration and hence
are different for the different thrust definitions
presented.\cite{ADS} As discussed
previously, all $x$ dependent logarithms have been resummed in the
parton density function $q(x,\tau Q^2)$ and the resulting form factor
is now $x$ independent.

One can expand this formula to ${\mathcal{O}}(\alpha_s^2)$ and compare
the results at small $\tau$ to those from fixed order Monte Carlo
programs. On doing so we find reasonable agreement with DISASTER++ but
disagreement with DISENT. The exact nature of this disagreement is
detailed in our paper.\cite{ADS}

Our result valid for small $\tau$ can then be matched to NLO
estimates from the Monte Carlo programs. The matching procedure extends the range of applicability
of the resummed calculation to larger $\tau$ values where non
logarithmic pieces in the fixed order result are significant.
The essential idea of matching is simple: one simply adds the resummed 
and NLO calculations and then subtracts the pieces corresponding to
double counting. Schematically we can write
\begin{equation}
R_{\mathrm{mat}} = R_{\mathrm{res}} + (R^{\mathrm{NLO}}-R_{\mathrm{res}}^{\mathrm{NLO}})
\end{equation}
In the above $R_{\mathrm{mat}}$ denotes the matched shape cross
section and $R_{\mathrm{res}}$ the resummed calculation, while the piece
 $(R^{\mathrm{NLO}}-R_{\mathrm{res}}^{\mathrm{NLO}})$ takes care of
 adding terms in the NLO result for $R$ that are absent in the
 resummed result expanded to the same order. In practice matching is a 
 technically involved procedure and different matching schemes can be
 used which differ on how to treat subleading logarithmic terms. 
Our prefered matching prescription is presented in the full
paper.\cite{ADS}

Once the resummation is carried out and matched to fixed order, we
have the best available perturbative prediction at hand. This
prediction is at least as good as the NLO Monte Carlo estimate over
the entire range of shape variable values and is far superior at small 
$\tau$. However we have not addressed another potential source of
uncertainty, namely power behaved corrections.

The main effect of power corrections will be to simply shift the
perturbative distributions by an amount proportional to $1/Q$ towards
larger $\tau$ values. An exception to this rule is the current jet
broadening variable where the perturbative prediction is squeezed and
shifted due to non-perturbative effects. The amount of the shift  
for the thrust variables is identical to the power correction to the
corresponding mean values which have been computed.\cite{self}

Other variables that are being studied at the moment include the
thrust defined wrt the actual thrust axis (normalised to $E$), the
current jet-mass and the $C$ parameter.\cite{DS} 
When resummed results for these distributions become available they
should also be 
matched to NLO and adjusted for power corrections after which detailed 
phenomenological analysis and comparisons with data should become possible.
\section*{Acknowledgments}
The work presented here was carried out in collaboration with Vito
Antonelli and Gavin Salam.

\end{document}